\documentclass[conference]{IEEEtran}
\IEEEoverridecommandlockouts
\usepackage{amsmath,amsfonts}
\usepackage[linesnumbered,ruled,vlined]{algorithm2e}
\usepackage{array}
\usepackage{subfigure}
\usepackage{textcomp}
\usepackage{stfloats}
\usepackage{url}
\usepackage{verbatim}
\usepackage{graphicx}
\usepackage{cite}
\usepackage{braket}
\usepackage{orcidlink}
\usepackage{diagbox}
\usepackage{makecell}
\usepackage {multirow}
\usepackage{fancyhdr}
\hypersetup{
colorlinks=true,
linkcolor=black,
citecolor=black,
urlcolor = black
}
\def\BibTeX{{\rm B\kern-.05em{\sc i\kern-.025em b}\kern-.08em
    T\kern-.1667em\lower.7ex\hbox{E}\kern-.125emX}}
\begin{document}

\title{Parallel Segment Entanglement Swapping\\
\thanks{The research leading to these results received funding from the National Key Research and Development Program of China under Grant 2023YFB2904005. (Corresponding author: Dong Zhang)}
}
\author{\IEEEauthorblockN{1\textsuperscript{st} Binjie He}
\IEEEauthorblockA{\textit{College of Computer and Data Science} \\
\textit{Fuzhou University}\\
Fuzhou, China \\
\textit{School of Information Technology}\\
\textit{Deakin University}\\
Melbourne, Australia\\
hebinjie33@gmail.com}
\and
\IEEEauthorblockN{2\textsuperscript{nd} Seng W. Loke}
\IEEEauthorblockA{\textit{School of Information Technology} \\
\textit{Deakin University}\\
Melbourne, Australia \\
seng.loke@deakin.edu.au}
\and
\IEEEauthorblockN{3\textsuperscript{rd} Dong Zhang}
\IEEEauthorblockA{\textit{College of Computer and Data Science} \\
\textit{Fuzhou University}\\
Fuzhou, China \\
\textit{Zhicheng College} \\
\textit{Fuzhou University}\\
Fuzhou, China \\
zhangdong@fzu.edu.cn}
}

\maketitle
\thispagestyle{fancy}
\fancyhead{}
\lhead{\footnotesize{This paper has been published in IEEE International Conference on Quantum Communications, Networking, and Computing (QCNC 2024), Kanazawa, Japan, 2024, pp. 271-279, doi: 10.1109/QCNC62729.2024.00050.}  \\ 
\footnotesize{©IEEE 2024. Personal use of this material is permitted. Permission from IEEE must be obtained for all other uses, in any current or future media including reprinting/republishing this material for advertising or promotional purposes, creating new collective works, for resale or redistribution to servers or lists, or reuse of any copyrighted component of this work in other works.}}
\cfoot{\quad}
\renewcommand{\headrulewidth}{0pt}
\pagestyle{empty}
\begin{abstract}
In the noisy intermediate-scale quantum era, scientists are trying to improve the entanglement swapping success rate by researching anti-noise technology on the physical level, thereby obtaining a higher generation rate of long-distance entanglement. However, we may improve the generation rate from another perspective, which is studying an efficient entanglement swapping strategy. This paper analyzes the challenges faced by existing entanglement swapping strategies, including the node allocation principle, time synchronization, and processing of entanglement swapping failure. We present Parallel Segment Entanglement Swapping (PSES) to solve these problems. The core idea of PSES is to segment the path and perform parallel entanglement swapping between segments to improve the generation rate of long-distance entanglement. We construct a tree-like model as the carrier of PSES and propose heuristic algorithms called Layer Greedy and Segment Greedy to transform the path into a tree-like model. Moreover, we realize the time synchronization and design the on-demand retransmission mechanism to process entanglement swapping failure. The experiments show that PSES performs superiorly to other entanglement swapping strategies, and the on-demand retransmission mechanism can reduce the average entanglement swapping time by 80\% and the average entanglement consumption by 80\%.
\end{abstract}


\section{Introduction}
Quantum network communication utilizes quantum teleportation to teleport qubits between two target quantum systems, but the precondition is that there must be quantum entanglement pairs between systems. In long-distance quantum communication, the noise in channels and devices will inevitably severely impact the entanglement preparation, entanglement distribution, and entanglement swapping, thereby reducing the efficiency of generating long-distance entanglement pairs. Hence, generating long-distance entanglement pairs is one of the toughest challenges in quantum communication. This paper focuses on the entanglement swapping process and presents an innovative strategy named Parallel Segment Entanglement Swapping (PSES) to improve the generation rate of long-distance entanglement.

Entanglement swapping (ES) means using a Bell State Measurement (BSM) for two qubits in two independent entangled pairs; the remaining two qubits go into an entangled state\cite{pan1998experimental}. Sequential ES\cite{coopmans2021netsquid, repeaterchainexample2021} is the most straightforward strategy in long-distance quantum communication. Repeaters perform ES one by one in sequential ES. Because only one repeater can perform ES each time, the generation rate of long-distance entanglement is low in sequential ES. In recent years, researchers have begun to consider parallel ES to improve the generation rate of entanglement; the main idea is to improve efficiency by making some nodes perform ES simultaneously. The existing research on parallel ES is divided into two categories: the Balanced Binary Tree (BBT) strategy\cite{briegel1998quantum, duan2001long, sangouard2011quantum, dai2020optimal} and the Imbalanced Binary Tree (IBT) strategy\cite{ghaderibaneh2022efficient}. 

In this paper, we analyze and summarize the problems and challenges faced by parallel ES (see Sec. \ref{chanllenges} for details), as briefly shown in the following.

\begin{itemize}
\item How do we allocate nodes reasonably for parallel ES? 
\item How do we solve the time synchronization problem of parallel ES? 
\item How do we deal with the ES failure? 
\item Either BBT or IBT has shortcomings, such as generating an ES tree without considering environmental interference and limiting the parent node to only a single repeater, which makes them unable to maximize efficiency.
\end{itemize}

To address these challenges of existing research, we propose the PSES strategy. The core idea of PSES is to segment nodes on the path, implementing the mechanism of ``performing parallel ES between segments and sequential ES within segments". We present our main contributions below.

(a) {\em Proposing the tree-like model, the definition and quantization formula of node ES time cost, and implementation algorithms of PSES.} Simulation results show that PSES's performance is superior to that of other parallel ES strategies.

(b) {\em Realizing the time synchronization mechanism.} With the help of the central controller of hierarchical architecture\cite{he2024hierarchical},  we design and deploy the time synchronization control program to ensure rounds of parallel ES work smoothly.

(c) {\em Proposing an on-demand retransmission mechanism for ES failure.} We utilize the central controller to identify the ES failed nodes in time and conduct targeted entanglement re-preparation. The experimental results show that the on-demand retransmission mechanism reduces both time cost and entanglement consumption by 80\% compared with the traditional full-path retransmission mechanism.

(d) Compared to anti-noise research\cite{kong2023entanglement, liu2022all, matsuo2018analysis} from the physical perspective, {\em PSES improves the generation rate of long-distance entanglement from the perspective of protocols.} PSES has a better generation rate of long-distance entanglement and lower resource cost than other ES strategies, which means that it can provide more long-distance entanglements for future distributed quantum computing, laying a foundation for implementing distributed quantum computing. Moreover, PSES ameliorates the network layer protocol of the hierarchical communication model\cite{he2024hierarchical}, which is one promising solution for the future quantum Internet. 

The rest of this paper is organized as follows. Sec. \ref{related_work} presents the related work. Sec. \ref{chanllenges} analyzes and summarizes the challenges faced by parallel ES. In Sec. \ref{designing_implementation}, we propose the PSES strategy. Sec. \ref{evaluation} shows the performance evaluation with simulation experiments. In Sec. \ref{conclusion}, we conclude this work.

\section{Related Work}
\label{related_work}
The work in\cite{sinclair2014spectral, dai2021entanglement} gave two opinions: a) the ES strategy should aim at improving efficiency because long-distance communication is sensitive to ES time; b) designing the ES strategy should carefully consider the dependencies between nodes.

Nodes on the path perform ES one by one in the sequential ES\cite{coopmans2021netsquid, repeaterchainexample2021}. However, Shchukin \emph{et al.} \cite{shchukin2019waiting} pointed out that in N-hop path ES, the post-order node must wait for the pre-order to complete the operation and transmit the BSM results before proceeding to the following action, and the waiting time is affected by the success rate of ES and the transmission rate of classical channels. The above research implied that long-distance communication would likely fail in a noisy environment due to the long waiting time of sequential ES.

To reduce waiting time and improve efficiency, researchers considered parallel ES strategies, which allow nodes on the path to perform ES simultaneously. The work in\cite{liu2017semihierarchical} pointed out that the constraint of parallel ES is that adjacent nodes cannot perform ES simultaneously because this means performing multiple quantum operations on a qubit simultaneously, which may lead to an error. Therefore, parallel ES can only be performed in non-adjacent nodes. Balanced Binary Tree strategy is a class of parallel ES. The BBT\cite{briegel1998quantum, duan2001long, sangouard2011quantum, dai2020optimal} strategy converts the path into a balanced binary tree, and the parent nodes of the same layer can perform ES simultaneously. The BDCZ protocol\cite{briegel1998quantum} performs parallel ES in multiple rounds. In each round, the path is divided into multiple segments of the same length, and the ES and entanglement purification are performed in parallel between the segments to ensure that the fidelity of long-distance entanglement is not too low. DLZC\cite{duan2001long, sangouard2011quantum} utilizes atomic ensemble and linear optics to enable the implementation of the core concepts of the BDCZ protocol. RED\cite{dai2020optimal} provides a mathematical method for forming an optimal BBT strategy under ideal environmental interference conditions. The above research shows that although the performance of the BBT strategy can be better than that of sequential ES, node environmental interference is not considered a factor in the strategy, so its performance may not always be satisfactory in the quantum network with node state differences. Aiming at the shortcomings of BBT, Ghaderibaneh  \emph{et al.}\cite{ghaderibaneh2022efficient} proposed an Imbalanced Binary Tree strategy. The core idea of IBT is to transform the path into an imbalanced binary tree according to nodes' ES efficiency to ensure that parent nodes in the same layer have similar time consumption, thereby reducing the waiting time and increasing the generation rate of long-distance entanglement. However, quantum devices' environmental interference, such as channel noise, dephasing noise, and depolarizing noise, may change over time, giving the node state a certain degree of randomness and volatility in the real world. Therefore, it may be difficult for IBT to find parallel nodes with similar ES time costs in the real world.

\section{Challenges and Motivations}
For clarity, we first briefly define node ES time cost here, which is detailed in Sec. \ref{node_evaluation}. The node ES time cost is the time required to complete a successful ES in a node. For example, if a node tries ES five times before it succeeds, the total time spent is denoted as node ES time cost.
\label{chanllenges}
\subsection{The Principle of Node Allocation}
The essence of parallel ES is to allow some nodes to perform ES simultaneously. After several rounds of operations, long-distance entanglement can be established between the sender and receiver. Reasonable allocations of nodes for each round are the key to improving the efficiency of parallel ES. The ideal principle would be to have parallel nodes take the same ES time cost for each round so there is no waiting time between nodes, thereby maximizing parallel efficiency.

The principle of node allocation in the existing studies\cite{briegel1998quantum, duan2001long, sangouard2011quantum, dai2020optimal,ghaderibaneh2022efficient} has some shortcomings. The path is only converted into a balanced binary tree by default without considering other factors in the BBT strategy, which makes its efficiency low. The IBT strategy considers the success rate of ES, but the channel noise and quantum memory efficiency are ignored, so that the node allocation is not reasonable enough. Therefore, establishing a reasonable node allocation principle is one of the problems faced by parallel ES. In this paper, we evaluate the node ES time cost by environmental interference and give the quantitative evaluation formula (see Sec. \ref{node_evaluation}).

\subsection{Time Synchronization}
\begin{figure*}[tp]
\centering
\includegraphics[width=6in]{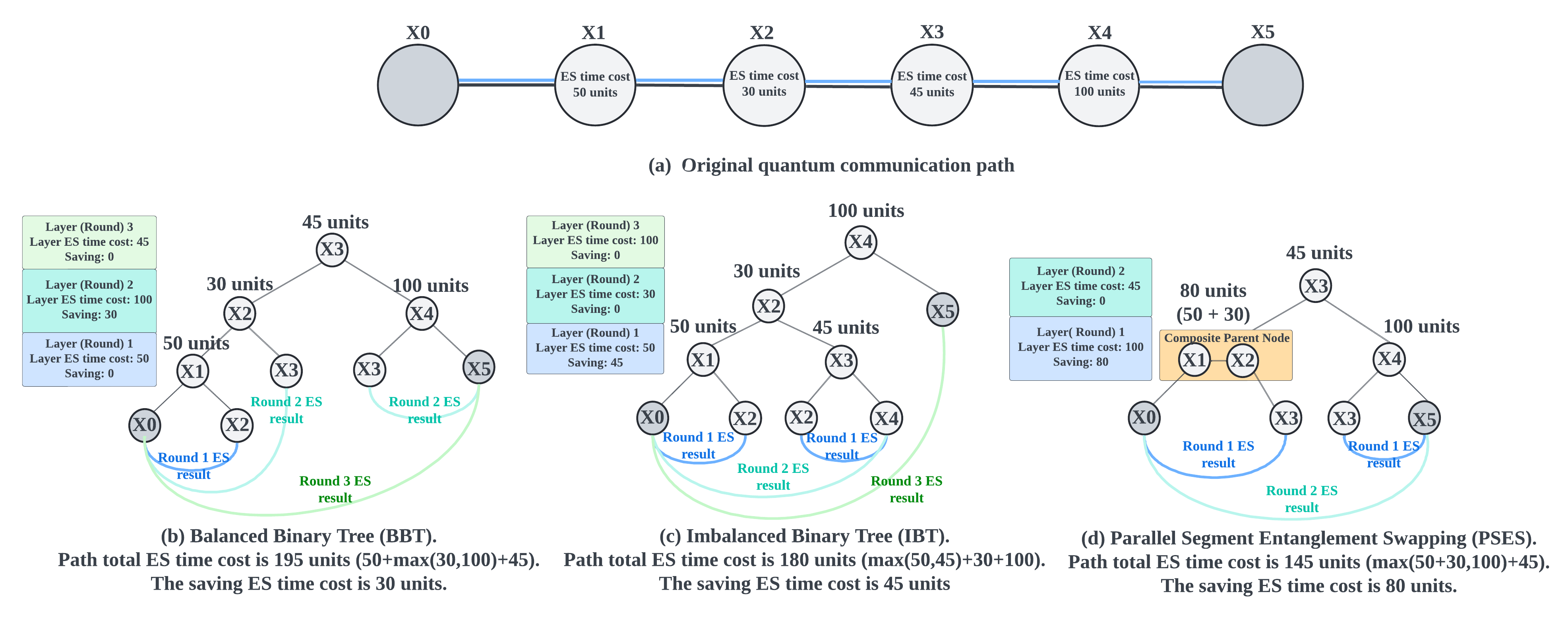}
\caption{Performing different parallel ES strategies on one path. (a) The original path consists of users (x0 and x5) and repeaters (x1, x2, x3, and x4). In the real world, the node ES time cost is mainly affected by the repeater's environmental interference (e.g., channel noise, dephasing noise, and depolarizing noise). (b) BBT transforms the path into balanced binary trees without considering node ES time cost, and the parent nodes of the same layer perform parallel ES to save time. (c) IBT transforms the path into imbalanced binary trees according to node ES time cost. (d) PSES introduces the composite parent node, consisting of several repeaters. PSES performs sequential ES inside the composite parent node and parallel ES between parent nodes of the same layer, which can fill the idle waiting time of parent nodes in the same layer for further savings.}
\label{parallel_swapping}
\end{figure*}

The parallel ES needs a time synchronization control to decide when to start the next round of parallelism. We give an example to clarify the importance of time synchronization. As shown in Fig. \ref{parallel_swapping}(c), when the operations of nodes x1 and x3 in the first layer (round) are completed, the ES of node x2 in the second layer can be performed. Since the operation of node x2 in the second layer depends on the entanglement pairs (x0-x2 and x2-x4) produced by the ES of nodes x1 and x3 in the first layer, if there is no time synchronization control, the node x2 in the second layer starts to execute in advance, the communication of the entire path will fail. So, time synchronization is crucial for parallel ES. However, the time synchronization problem is not discussed in the existing research\cite{briegel1998quantum, duan2001long, sangouard2011quantum, dai2020optimal, ghaderibaneh2022efficient}. In this research, we use the advantages of hierarchical architecture to design a time synchronization control module for parallel ES (see Sec. \ref{time_sync_swapping_failure}).

\subsection{Entanglement Swapping Failure}
\label{entanglement_swapping_failure}
Due to qubits being particularly sensitive to environmental interference, ES failures are very common in quantum communication. In parallel ES, the ES failure of a node will hinder path communication. As shown in Fig. \ref{parallel_swapping}(c), when the ES of the node x1 of the first layer (round) fails, the entanglement between nodes x0 and x2 cannot be obtained, and then the node x2 of the second layer has no conditions to perform operations. So, ES failure is an urgent problem in parallel ES strategy.

The way to deal with node ES failures in the existing research\cite{briegel1998quantum, duan2001long, sangouard2011quantum, dai2020optimal, ghaderibaneh2022efficient} can be called full-path retransmission, which is to re-prepare and distribute entanglements on the full path and perform parallel ES from the beginning. It is obvious that full-path retransmission will greatly increase the delay in long-distance quantum communication because the more nodes on the path, the greater the probability of node ES failure, resulting in an increase in the number of full-path retransmissions and the consumption of time. With the help of hierarchical architecture\cite{he2024hierarchical}, we propose a more efficient on-demand retransmission mechanism for ES failures (see Sec. \ref{time_sync_swapping_failure}).

\subsection{The Analysis of Existing Parallel Entanglement Swapping Strategies}
Suppose we have a quantum communication path, as shown in Fig. \ref{parallel_swapping}(a). When we perform the sequential ES strategy on this path, the path total ES time cost is the sum of all node ES time costs, that is, $50+30+45+100=225$ units. In the following part, we use the case in Fig. \ref{parallel_swapping} to analyze the ES time saved by different parallel ES strategies compared with sequential ES strategy.

The core idea of the BBT\cite{briegel1998quantum, duan2001long, sangouard2011quantum, dai2020optimal}  is to transform the path into a balanced binary tree and simultaneously perform ES between the parent nodes in the same layer (round) to establish the entanglement between the child nodes. For example, the BBT transforms the path (see Fig. \ref{parallel_swapping}(a)) into a tree (see Fig. \ref{parallel_swapping}(b)). The first layer performs the ES in node x1 to establish entanglement between nodes x0 and x2, and the second layer performs the ES in nodes x2 and x4 in parallel to establish entanglements between nodes x0 and x3 as well as nodes x3 and x5, respectively. The subsequent layers can be done in the same manner. In this case, BBT utilizes the parallelism of nodes x2 and x4 to save 30 units of ES time costs. Specifically, for the first layer, only node x1 with 50 units of ES time cost performs ES, so the ES time cost of the first layer is 50 units. For the second layer, parallel ES is performed between node x2 with 30 units of ES time cost and node x4 with 100 units of ES time cost, so the ES time cost of the second layer is equal to the max ES time cost of the node in this layer, that is, $\max(30,100)=100$ units. For the third layer, only node x3 with 45 units of ES time cost performs ES, so the ES time cost of the third layer is 45 units. The path total ES time cost is equal to the sum of all layer ES time costs, so the total ES time cost of the path is $50+\max (30,100) +45=195$ units. BBT saves $225-195=30$ units of ES time cost compared to the sequential ES strategy in this case. However, the disadvantage of BBT is that it only requires the tree to be balanced without considering node ES time cost, thereby limiting the efficiency of its improvement. We can see this shortcoming by comparing BBT with other strategies in Fig. \ref{parallel_swapping}.

The core idea of the IBT\cite{ghaderibaneh2022efficient} is to transform the path into an imbalanced binary tree according to the node ES time cost so that the ES time costs of parent nodes in each layer are as close as possible to maximize efficiency. In Fig. \ref{parallel_swapping}(c), IBT performs parallel ES in nodes x1 and x3 in the first layer, establishing entanglements between nodes x0 and x2 as well as nodes x2 and x4, respectively. The subsequent layers can be done in the same manner. The path total ES time cost of the IBT is 180 units, less than the BBT, which can be calculated the same way as the BBT. However, in the real world, the ES time cost of a single quantum repeater is mainly determined by environmental interference, which may change over time, thereby giving the node ES time cost a certain degree of randomness and volatility. Thus, it may be hard for IBT to find nodes with similar ES time costs for parallel ES. Therefore, the disadvantage of the IBT is that it limits the parent node to consisting only of a single node, which makes its efficiency less ideal in reality than in theory. We can see this shortcoming by comparing IBT with our proposed PSES (see Fig. \ref{parallel_swapping}(d)).

\section{The Design and Implementation of Parallel Segment Entanglement Swapping}
\label{designing_implementation}
\subsection{Tree-like Model Design}
\label{tree_like_model}
PSES transforms the path into a tree-like model with N layers (rounds), and each layer is formed according to the principle of segmentation. For example, in the first layer of Fig. \ref{parallel_swapping}(d), the original path of Fig. \ref{parallel_swapping}(a) is divided into two segments, \{x0,x1,x2,x3\} and \{x3,x4,x5\}. The two nodes at the head and tail of the segment are the leftmost leaf node and the rightmost leaf node, respectively, and the middle part of the segment constitutes the composite parent node. For a segment, the parent node performing ES establishes an entanglement between the leftmost leaf node (segment head) and the rightmost leaf node (segment tail). For example, in the segment \{x0,x1,x2,x3\}, x1 and x2 perform ES to establish an entanglement between x0 and x3. For the n+1 layer, it is to segment the remaining path of the n layer. For example, if ES is successful in the first layer of Fig. \ref{parallel_swapping}(d), the remaining path will be x0-x3-x5, so the segmentation of the second layer is \{x0,x3,x5\}, where x3 is its parent node, x0 is the leftmost leaf node, and x5 is its rightmost leaf node.

The node ES time cost is quantitatively evaluated according to the dephasing rate, depolarizing rate, and Q-channel quality. Segmentation is performed according to the node ES time cost to maximize the saving time.

The principle of ``performing parallel ES between segments and sequential ES within segments" is applied when executing the tree-like model. For example, in the first layer of Fig. \ref{parallel_swapping}(d), two composite parent nodes \{x1, x2\} and \{x4\}, which belong to two segments \{x0, x1, x2, x3\} and \{x3, x4, x5\}, begin to perform ES simultaneously. However, inside the composite parent, the nodes x1 and x2 perform sequential ES.

Even though Fig. \ref{parallel_swapping} is a simple case, there is no loss of generality. PSES allows more parallelism to be exploited than BBT or IBT as long as nodes in any path can be composed into composite parent nodes.

\subsection{The Definition and Quantitative Evaluation of Node Entanglement Swapping Time Cost}
\label{node_evaluation}
{\bfseries Definition:} The node ES time cost is the time cost of completing a successful ES in a node. For example, if a node retries ES five times before it succeeds, the total time spent is denoted as node ES time cost. Node ES time cost reflects the relative time length instead of the precise time because we adopt environmental interference to evaluate node ES time cost. For example, we can say that 2 units of node ES time cost is longer than 1 unit, but we cannot say that 2 units represent some precise time, such as 2 ms or 2 s. Because distributing entanglement via the quantum channel and storing entanglement in the quantum memory are required before ES, the environmental interference, including quantum channel noise, depolarizing noise, and dephasing noise, affects the number of retries of the node to get a successful ES, thus determines the node ES time cost. The quantum channel noise affects the process of entanglement distribution, thus affecting the ES. The depolarizing noise affects the fidelity of entanglement in quantum memory. The dephasing noise introduced by quantum operation affects the fidelity of the new entanglement produced by ES. Both dephasing noise and dephasing noise will affect the ES success rate because entanglement can not be used in ES or other quantum operations when the fidelity of entanglement is less than 0.5 \cite{kozlowski2023rfc}. In other words, if the fidelity of the new entanglement produced by ES is less than 0.5, the ES fails. We should note that this research supposes the quantum network has a constant entanglement preparation rate, the same implemented physical system (e.g., trapped-ion or superconducting), and two quantum memories in each node (i.e., only one ES allowed per time). All the factors above may affect ES and are worth studying in the future, but this study focuses on the influence of environmental interference on ES.
 
{\bfseries Quantitative Evaluation:} In a simulated environment (e.g., NetSquid \cite{coopmans2021netsquid}), the dephasing noise is represented by the dephasing rate, which affects the ES success rate; the depolarizing noise is represented by the depolarizing rate, which affects the quantum memory efficiency; and the Q-channel quality is represented by the Q-channel loss init rate and Q-channel loss noise which affect the entanglement distribution. The successful execution of ES in the repeater needs to meet three conditions: a) entanglement distribution is successful; b) qubits are successfully stored in the quantum memory; and c) ES operation succeeds. Therefore, we propose formula (\ref{node_cost}) to evaluate node ES time cost quantitatively. 
\begin{equation}
\label{node_cost}
NC = \frac{1}{(1-DPZR)\times(1-DPSR)\times CQ}
\end{equation}
NC stands for node ES time cost, DPZR for the depolarizing rate, DPSR for the dephasing rate, and CQ for channel quality. In formula (\ref{node_cost}), (1 - DPZR) represents the probability of a qubit's successful storage in memory without depolarization; (1-DPSR) represents the probability of successful operation on ES; and CQ represents the probability of a qubit's successful distribution via the quantum channel. The denominator in formula (\ref{node_cost}) represents the probability that the repeater completes one ES, so its reciprocal represents the number of attempts required by the repeater to complete one ES, which is the node ES time cost. In an ideal quantum network without environmental interference, the depolarizing rate and dephasing rate are both 0, the channel quality is 1, and the node ES time cost is 1, representing the need for one-time ES to succeed. When the environmental interference increases, the node state worsens, and the number of attempts increases, which means the node ES time cost increases.

According to the research \cite{he2024hierarchical}, when the noise of a 100 km channel is greater than or equal to 0.2 dB/km, the success rate of entanglement distribution is zero, so the maximum acceptable noise of a 100 km channel is 0.2 dB/km. Therefore, if we assume the length of a quantum channel is 100 km, the channel quality can be expressed by the formula (\ref{channel_quality}). We note that the assumption is reasonable for facilitating experimental verification because we can control the channel length in the simulation. We can adjust the channel quality evaluation formula if there are other channel lengths.
\begin{equation}
\label{channel_quality}
CQ = 1-\frac{QLIR+\frac{QLN}{0.2}}{2}
\end{equation}
QLIR is for Q-channel loss init rate, and QLN is for Q-channel loss noise. Under the condition of our assumption, it has $0 \leq QLIR \leq 1$ and $0 \leq QLN \leq 0.2$. The derivation and correctness of formula (\ref{channel_quality}) have been proven in the research \cite{he2024hierarchical}.

\subsection{Implementation Algorithms}
\label{implementation_algorithms}
The primary problem PSES faces is finding an algorithm to transform paths into tree-like models. We have two possible approaches: an optimization approach (e.g., Dynamic Programming) and a heuristic approach (e.g., Greedy). Researchers\cite{ghaderibaneh2022efficient} pointed out that the time complexity of Dynamic Programming is enormous. Thus, we propose two feasible heuristic algorithms: Layer Greedy and Segment Greedy. Before describing algorithms, we should note that the tree-like model is generated from the bottom to the top layer. Hence, the solution chosen at the lower layer affects the higher layer.

{\bfseries Layer Greedy:} The core idea of Layer Greedy is to take the optimal solution of each layer as the greedy goal. The solution that saves the most ES time cost at the current layer is selected without considering the higher layer situation. The pseudocode for Layer Greedy is shown in Alg. \ref{algo_LG}. The Layer Greedy algorithm is divided into two steps, as shown below. 

Step A) For the first layer, all possible solutions of the original path are traversed, for example, \{\{x0,x1,x2,x3\},\{x3,x4,x5\}\} and \{\{x0,x1,x2\},\{x2,x3,x4,x5\}\} are two of the possible first layer solutions of the path in Fig. \ref{parallel_swapping}(a). The solution with the greatest ES time cost saving is selected as the first layer of the tree-like model. The remaining path consists of the nodes not selected as the parent node.

Step B) For the nth layer (n$>$1), recursively perform step A on the remaining path of the (n-1)th layer until the root.

 \IncMargin{1em}
\begin{algorithm}\SetKwInOut{Inputs}{inputs}\SetKwInOut{Output}{output}\SetKw{Break}{break} \SetKwFunction{LayerGreedy}{LayerGreedy}\SetKwProg{Fn}{Function}{:}{} \SetKwFunction{FindLayerSolutions}{FindLayerSolutions}
\let\oldnl\nl 
{\fontsize{10pt}{15pt}\selectfont
\newcommand{\nonl}{\renewcommand{\nl}{\let\nl\oldnl}}	
	\Inputs{path,costs} 
	\Output{final\_solution}
	\tcp{ls: an array of layer solutions}
	\tcp{lp: an array of left path}
	\tcp{cs: an array of cost saving}
	\tcp{sn: the number of segments}
 	 \Fn{\LayerGreedy{path,costs}}{
	 \While{path\_length $>$ 2}{
          ls,lp,cs = FindLayerSolutions(path,costs); \\
          best\_ls = ls[cs.index(max(cs))];\\
	 final\_solution.append(best\_ls);\\
	 path = lp[cs.index(max(cs))];\\
        \KwRet\ LayerGreedy(path,costs);
        }
        \KwRet\ final\_solution;
 	 }
	  \Fn{\FindLayerSolutions{path,costs}}{
	  max\_sn = RoundUp((path\_length-2)/2); \\
	  sn = 1;\\
	\While{sn $\leq$ max\_sn}{
	  \tcp{find all layer solutions from case sn is 1 to max\_sn.}
	  ls,lp,cs $\leftarrow$ TraversePath(path,costs,sn);\\
	  sn = sn+1;\\
	  }
	  \KwRet\ ls,lp,cs;
	  }
\caption{Layer Greedy.}
\label{algo_LG} 
}
\end{algorithm}
\DecMargin{1em} 

 \IncMargin{1em}
\begin{algorithm}\SetKwInOut{Inputs}{inputs}\SetKwInOut{Output}{output}\SetKw{Break}{break} \SetKwFunction{SegmentGreedy}{SegmentGreedy}\SetKwProg{Fn}{Function}{:}{}
\let\oldnl\nl 
{\fontsize{10pt}{15pt}\selectfont
\newcommand{\nonl}{\renewcommand{\nl}{\let\nl\oldnl}}	
	\Inputs{path,costs} 
	\Output{final\_solution}
	\tcp{fs: the first segment}
	\tcp{lp: an array of left path}
	\tcp{cs: the current segment}
 	 \Fn{\SegmentGreedy{path,costs}}{
	 \While{path\_length $>$ 2}{
          fs = [path[0],path[1],path[2]]; \\
          solution.append(fs);\\
          lp = path.remove(fs.parent\_node);\\
          MSC = FindCost(costs,fs);\\
	  i = 3; \tcp{node index}
	  cs = [];\\
	 \While{i in range(3,path\_length-1)}{
	 \If{len(cs) == 0}{
	 NBCS = min(costs[i],MSC) - max(0,costs[i]-MSC);\\
	 \If{NBCS$<$0}{
	 i++;\\}
	 \Else{
	 cs = [path[i-1],path[i],path[i+1]];\\
	 lp = path.remove(cs.parent\_node);\\
	 CCS = FindCost(costs,cs); i++;\\
	 \If{i == path\_length-1}{
	 solution.append(cs);\\
	 }
	 }
	 }
	 \Else{
	 NBFS = min(CCS+cost[i],MSC)-max(0,CCS+cost[i]-MSC);\\
	 NBG = NBFS-NBCS;\\
	 \If{NBG$<$0}{
	 solution.append(cs);\\
	 MSC = max(CCS,MSC);\\
	 cs = []; i++;\\
	 }
	 \Else{
	 cs.append(path[i+1]);\\
	 lp = path.remove(cs.parent\_node);\\
	 CCS = FindCost(costs,cs); i++;\\
	 \If{i==path\_length-1}{
	 solution.append(cs);\\
	 } 
	 }
	 }
	 }
	final\_solution.append(solution);\\
        \KwRet\ SegmentGreedy(lp,costs);
        }
        \KwRet\ final\_solution;
 	 }
\caption{Segment Greedy.}
\label{algo_SG} 
}
\end{algorithm}
\DecMargin{1em} 

{\bfseries Segment Greedy:} The core idea of Segment Greedy is to take the optimal segmentation solution as the greedy goal. When considering whether the current node can form a segment, the benefit of forming a segment is obtained through evaluation formulae. The decision is made to maximize the benefit. 

In PSES, there are two kinds of segment: segment with a single parent node and segment with a composite parent node. We present formula (\ref{current_segment_net_benefit}) to evaluate the benefit of the segment with a single parent node.  We refer to the currently computed segment as the current segment.
\begin{equation}
\begin{aligned}
\label{current_segment_net_benefit}
NBCS = \min(CCS,MSC) - \\ \max(0,CCS-MSC) 
\end{aligned}
\end{equation}
NBCS stands for the net benefit of the current segment, CCS for the ES time cost of the current segment, and MSC for the maximum segment ES time cost in the current layer. We note that the maximum segment ES time cost determines the total ES time cost of a layer. Formula (\ref{current_segment_net_benefit}) represents the net benefit of forming a new segment when the current traversing node acts as a single parent node. If the current segment ES time cost is less than the maximum segment ES time cost in the existing segments, the net benefit is the current segment ES time cost. If the current segment ES time cost exceeds the maximum segment ES time cost, the net benefit is the maximum segment ES time cost minus the difference between the current and maximum segment ES time costs. If the net benefit is greater than 0, we form the current segment.

We present formulae (\ref{future_segment_net_benefit}) and (\ref{net_benefit_growth}) to evaluate the benefit of the segment with a composite parent node. We refer to the currently computed segment as the future segment to distinguish it from the segment with a single parent node.
\begin{equation}
\begin{aligned}
\label{future_segment_net_benefit}
NBFS = \min(CCS+CN,MSC) - \\ \max(0,CCS+CN-MSC) 
\end{aligned}
\end{equation}
\begin{equation}
\label{net_benefit_growth}
NBG = NBFS - NBCS 
\end{equation}
NBFS stands for the net benefit of the future segment, CN for the ES time cost of the node, and NBG for the net benefit growth. Formulae (\ref{future_segment_net_benefit}) and (\ref{net_benefit_growth}) represent the net benefit growth of the current traversal node, which joins the parent node of the current segment to form a future segment. Formula (\ref{future_segment_net_benefit}) can be used to calculate the net benefit of the future segment with a composite parent node; its principle is the same as that of formula (\ref{current_segment_net_benefit}). Formula (\ref{net_benefit_growth}) shows the net benefit growth, which is the difference in net benefits between the future and current segments. If the net benefit growth is positive, we form the future segment.

Alg. \ref{algo_SG} shows the pseudocode for Segment Greedy. Segment Greedy forms the first segment with the first three nodes by default, assuming the path includes at least three nodes. Then, Segment Greedy traverses the path from the fourth node. If the traversed node is not a segment tail, it is decided whether to generate a new segment with a single parent node based on the net benefit of formula (\ref{current_segment_net_benefit}). If the traversed node is a segment tail, it is decided whether to join the parent node of the current segment to generate a new segment with a composite parent based on the net benefit growth of formulae (\ref{future_segment_net_benefit}) and (\ref{net_benefit_growth}). Each iteration of the outermost loop of Algorithm \ref{algo_SG} forms a layer of the tree-like model. The steps of Segment Greedy are shown below.

Step A) Traversing the original path for one time, making a greedy decision on whether each node is segmented according to evaluation formulae (net benefit or net benefit growth), and forming the remaining path with nodes not included in the parent nodes. 

Step B) Performing Step A recursively until reaching the root node.

\subsection{Time Synchronization and Entanglement Swapping Failure Processing}
\label{time_sync_swapping_failure}
\begin{figure}[tp]
\centering
\includegraphics[width=3in]{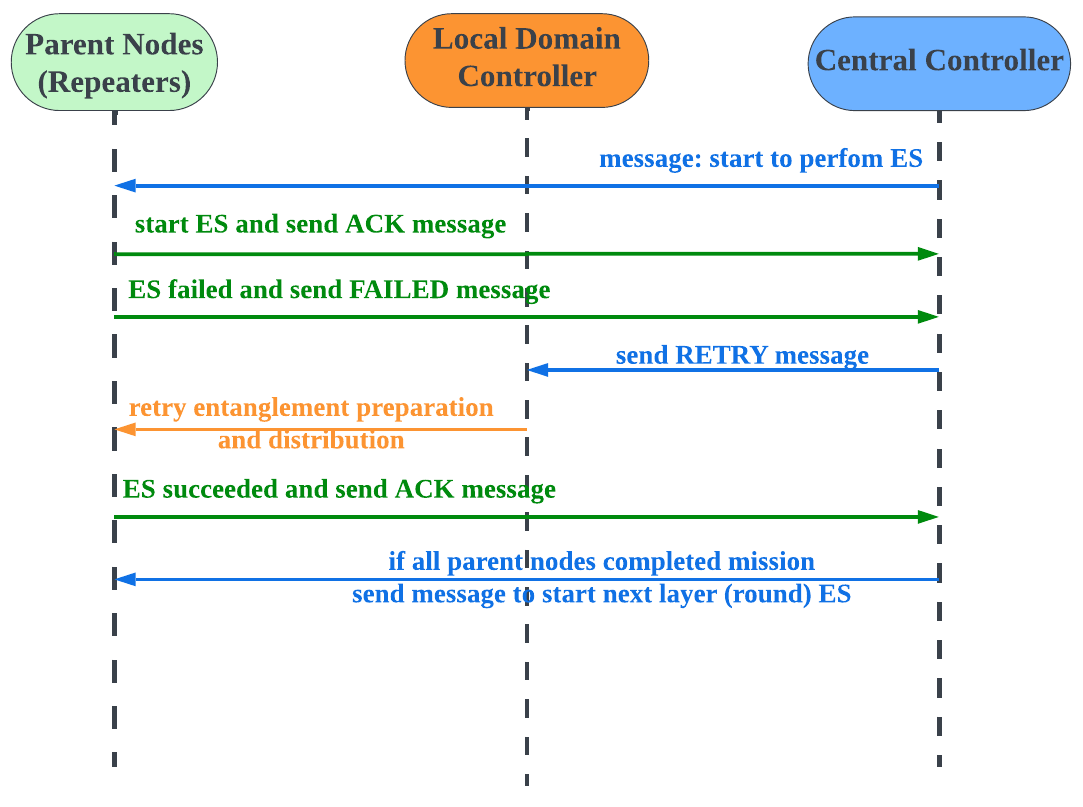}
\caption{The process of time synchronization and on-demand retransmission mechanism}
\label{time_sync_node_failure}
\end{figure}

\begin{figure}[tp]
\centering
\includegraphics[width=3in]{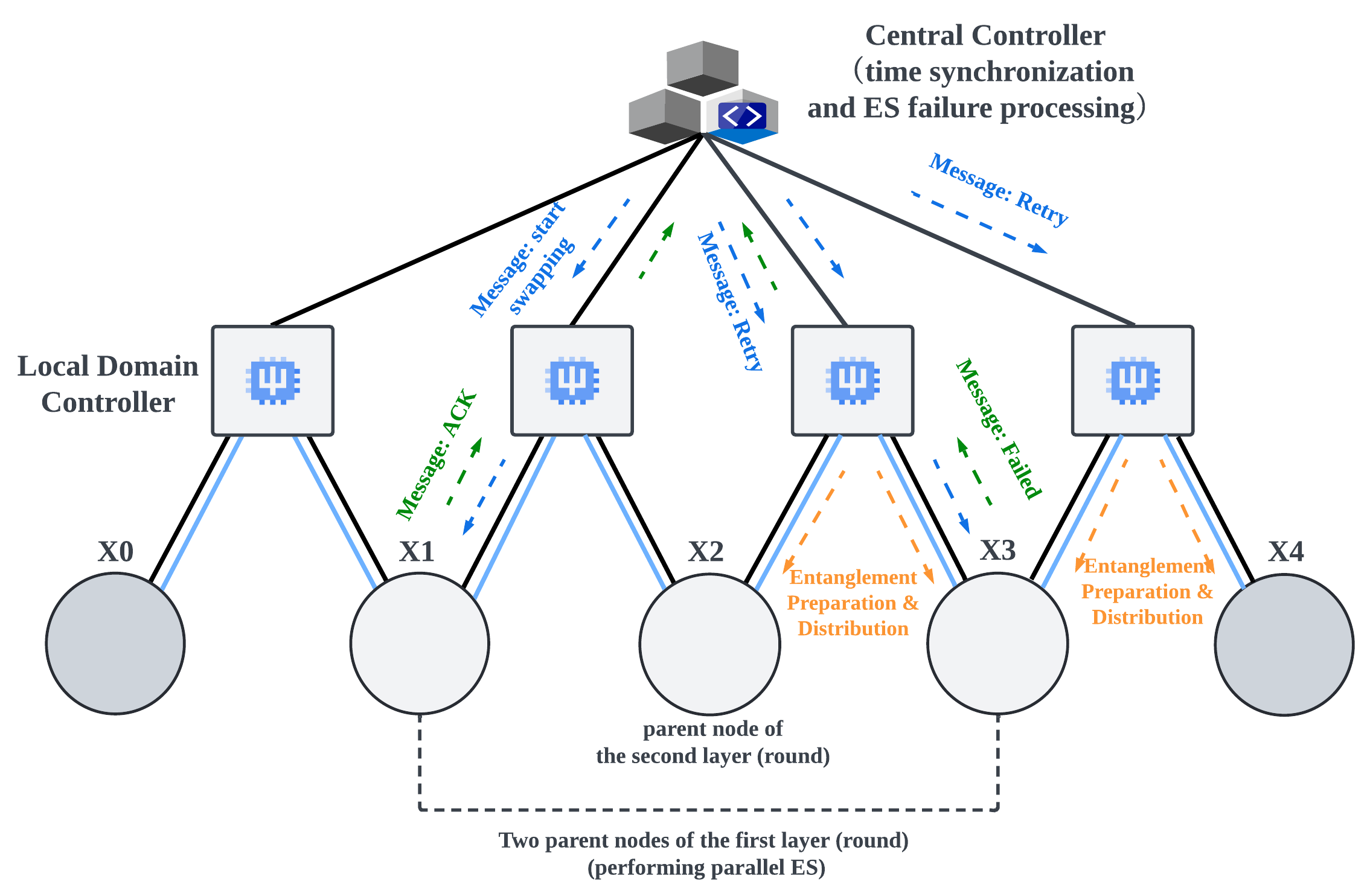}
\caption{The message flow of time synchronization and on-demand retransmission mechanism}
\label{message_flow}
\end{figure}

The key to realizing time synchronization is having a control device that can coordinate the process of parallel ES. The control device must communicate with repeaters and obtain the ES result in a timely manner. As shown in Fig. \ref{message_flow}, the central controller of hierarchical architecture\cite{he2024hierarchical} can communicate with repeaters through classical channels, making it a control device for parallel ES time synchronization. We deploy a parallel ES control program in the central controller, including the time synchronization function. Fig. \ref{time_sync_node_failure} and \ref{message_flow} show the core process. The central controller sends instructions to the parent nodes in the same layer before each parallel ES round begins. After receiving the instruction, the repeater synchronously performs the ES and sends an ACK message to the central controller. Once ES is completed, it sends another ACK message to the central controller. When receiving ACK messages from all parent nodes, the central controller confirms that all repeaters in the current round have completed the task and then starts to notify the parent nodes in the next layer for parallel ES. We should note that because the internal process of the composite parent node is sequential ES, the central controller starts notifying the posterior repeater to perform ES only after the previous repeater completes the ES. It ensures that the sequential ES inside the composite parent node can be successfully carried out.

We propose the on-demand retransmission mechanism for ES failure, which uses the central controller to identify the nodes that fail to perform ES and re-prepare and distribute entanglements in time for the failed nodes. The on-demand retransmission mechanism avoids re-preparing and distributing entanglement on the entire path due to the ES failure of some nodes, thereby reducing quantum communication time and increasing efficiency. Fig. \ref{time_sync_node_failure} and \ref{message_flow} show the core process. During a parallel ES round, the central controller receives a FAILED message if a parent node experiences a ES failure. Once the central controller receives this information, it will send a RETRY message to the corresponding local domain controller to re-prepare and distribute entanglement for related repeaters. The parent node immediately restarts the ES after receiving the new entanglement. We should emphasize that because repeaters inside the composite parent node are interdependent, when one of them experiences a ES failure, all related repeaters need to re-prepare and distribute entanglement.

\section{Simulation and Evaluation}
\label{evaluation}

\subsection{Environmental Setting and Methodology}
\begin{figure*}[tp]
\centering
\includegraphics[width=6in]{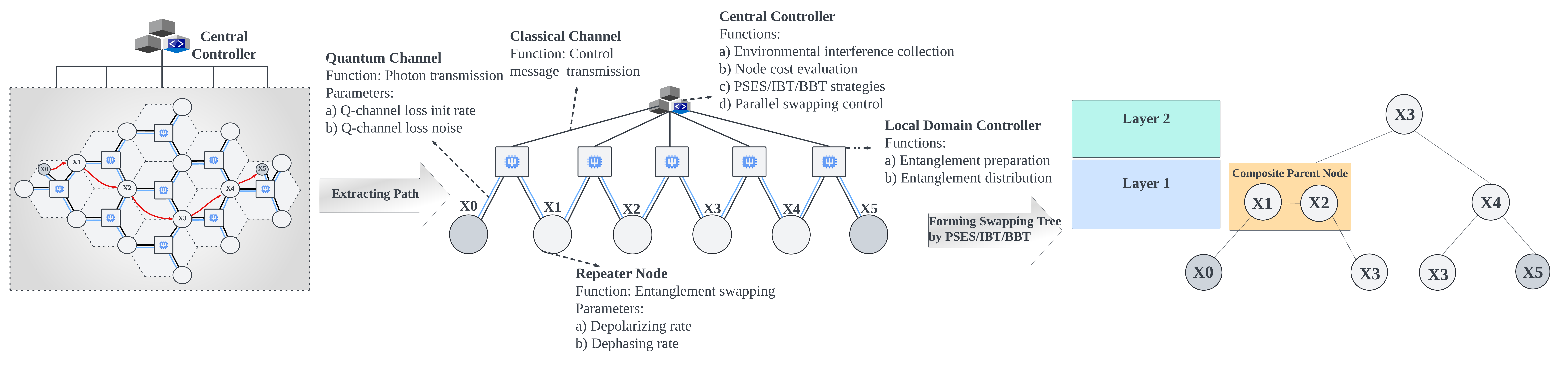}
\caption{Environmental setting of experiments. The end-to-end path is extracted from the cellular topology of hierarchical architecture. A path includes the central controller, local domain controllers, classical and quantum channels, and nodes. Parallel ES strategies are deployed in the central controller to transform the path into a swapping tree.}
\label{simulation_environment}
\end{figure*}

{\bfseries Environmental Setting:} We utilize NetSquid\cite{coopmans2021netsquid} to simulate a quantum network for evaluating PSES. The source code is published in\cite{pses2024}. Fig. \ref{simulation_environment} shows the experimental environment. We build a cellular topology of the hierarchical quantum network\cite{he2024hierarchical}. The cellular topology consists of a central controller and several domains (regular hexagonal regions in Fig. \ref{simulation_environment}). Because three repeaters are enough to cover a domain and allow regular hexagonal areas to be stitched together seamlessly\cite{he2024hierarchical}, each domain consists of a local domain controller, three repeaters, and several users. The local domain controller is responsible for the entanglement preparation and distribution of all devices in the domain. Any two quantum users in the cellular topology can perform quantum communication using the communication model provided by\cite{he2024hierarchical}. The component functions of the communication path are shown in Fig. \ref{simulation_environment}. After entanglement routing confirms a path, a tree-like model is formed by a parallel ES strategy and stored in the central controller waiting for execution.

{\bfseries Methodology:} We use the average ES time as the performance metric because the shorter the average ES time, the higher the generation rate of long-distance entanglement. It should be noted that the entanglement is invalid when the fidelity is less than 0.5\cite{kozlowski2023rfc}. Therefore, only the fidelity of long-distance entanglement greater than 0.5 can be counted as a successful ES in our experiments.

\subsection{Result}
\begin{figure}[tp]
\centering
\subfigure[]{
\centering
\includegraphics[width=1.6in]{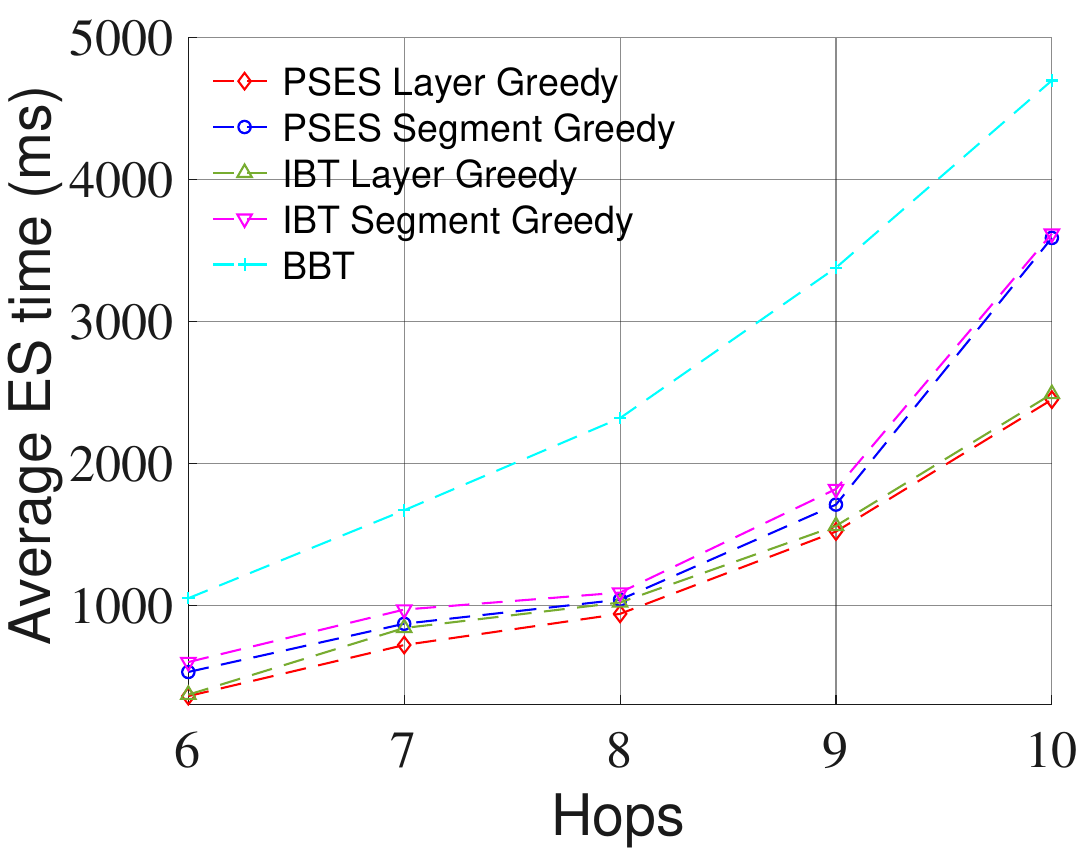}
}
\subfigure[]{
\centering
\includegraphics[width=1.5in]{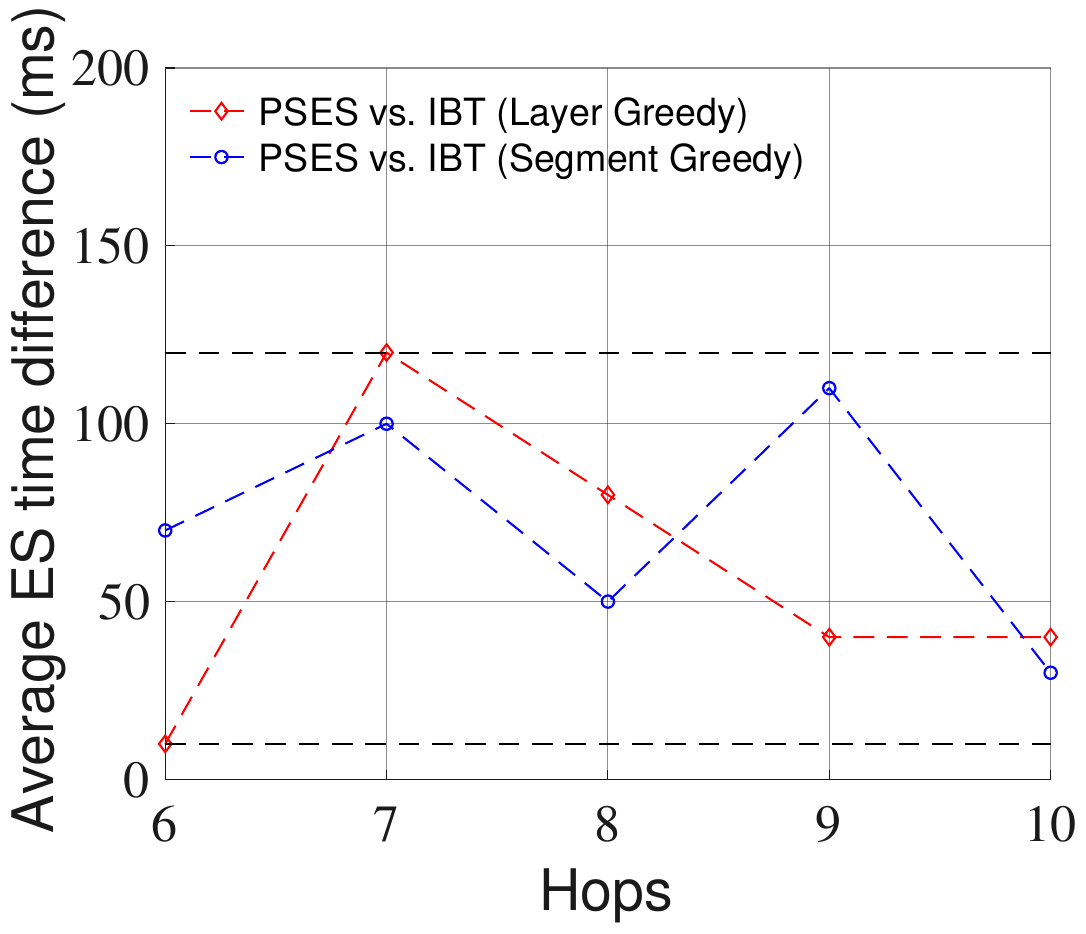}
}
\caption{Average ES time vs. hops. The average node ES time cost is 1.4 units. The standard deviation of node ES time cost is 0.1 units.}
\label{swapping_time_vs_hops}
\end{figure}

{\bfseries Average ES time vs. hops}: Fig. \ref{swapping_time_vs_hops}(a) shows that the average ES time of all parallel ES strategies increases with the hops change from 6 to 10. This phenomenon is reasonable because an increase in hops means that more repeaters must perform ES, so that the average ES time will also increase. Moreover, under the path with different hops, the average ES time always has BBT $>$ IBT Segment Greedy $>$ PSES Segment Greedy $>$ IBT Layer Greedy $>$ PSES Layer Greedy. Fig. \ref{swapping_time_vs_hops}(b) shows that the difference in average ES time between PSES and IBT does not increase with hop changes, indicating that hops will not impact the performance difference between PSES and IBT.

\begin{figure}[tp]
\centering
\subfigure[]{
\centering
\includegraphics[width=1.6in]{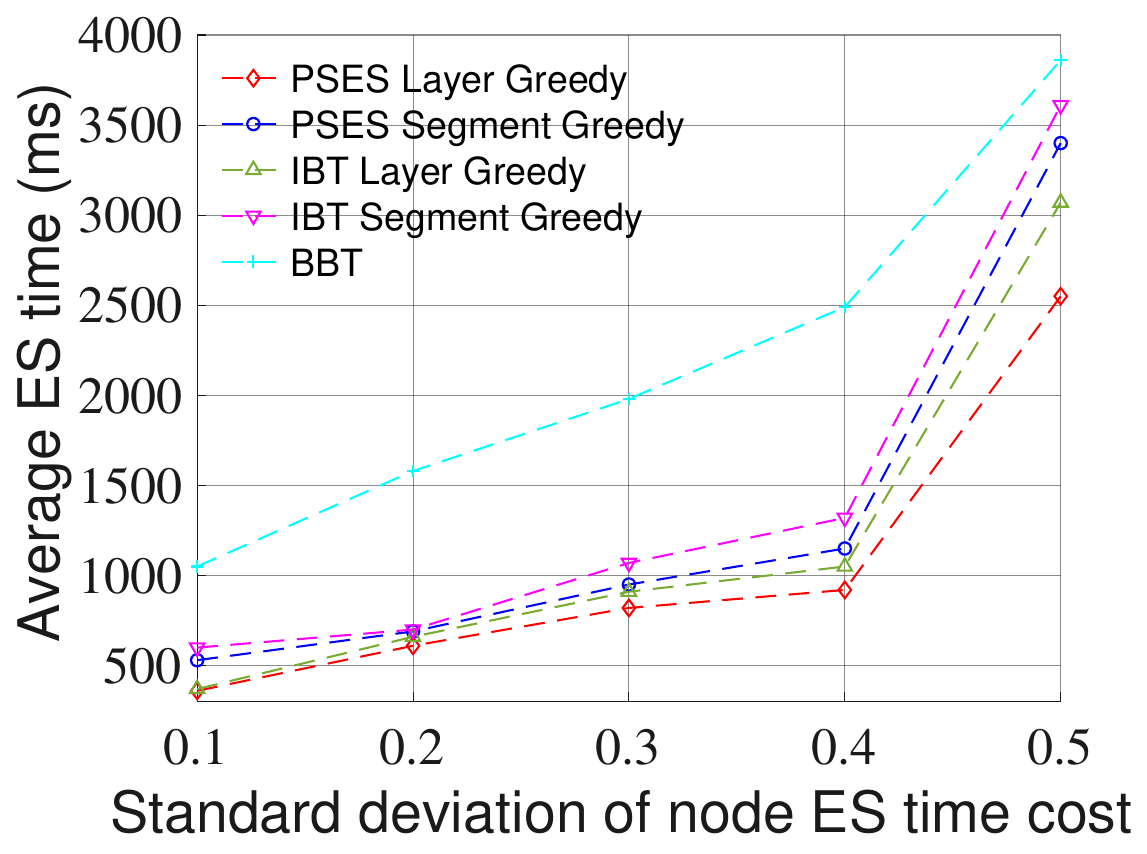}
}
\subfigure[]{
\centering
\includegraphics[width=1.6in]{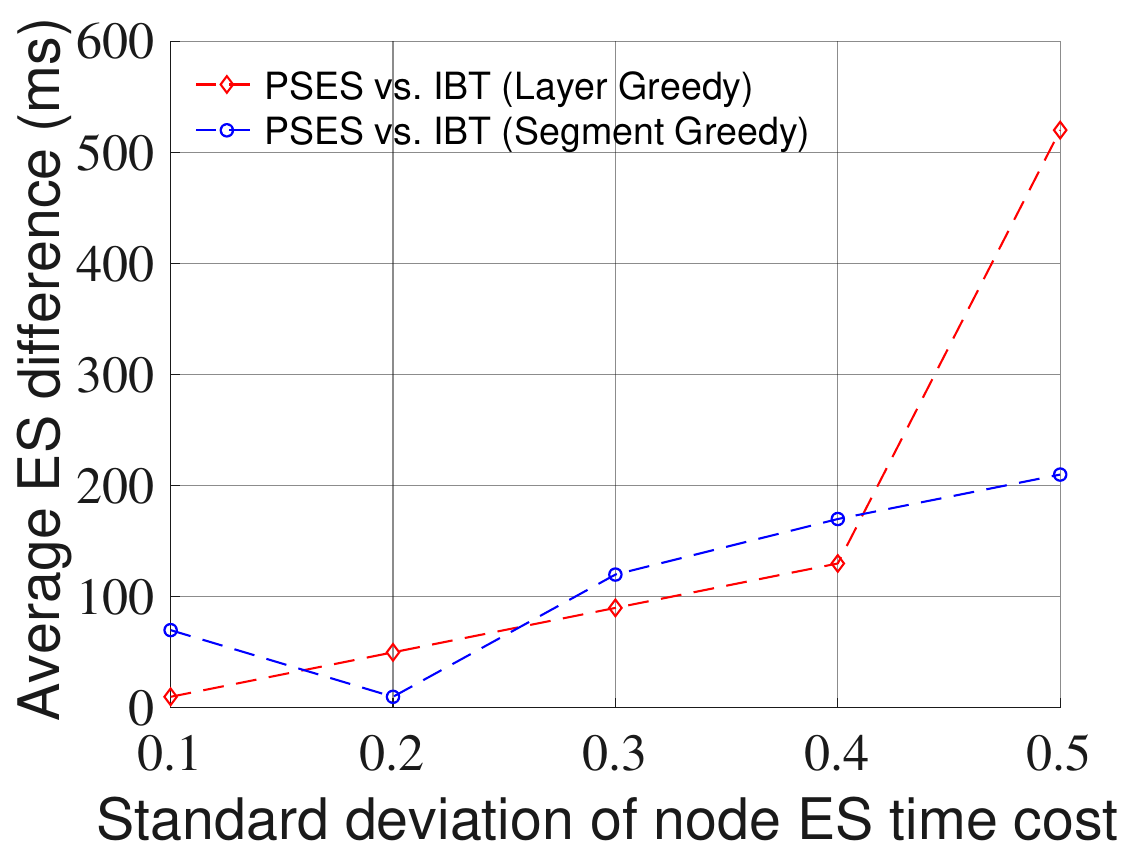}
}
\caption{Average ES time vs. standard deviation of node ES time cost. The average node ES time cost is 1.4 units. The path hops are 6.}
\label{swapping_time_vs_std}
\end{figure}

\begin{figure}[tp]
\centering
\begin{minipage}[t]{0.49\linewidth}
\centering
\includegraphics[width=1.8in]{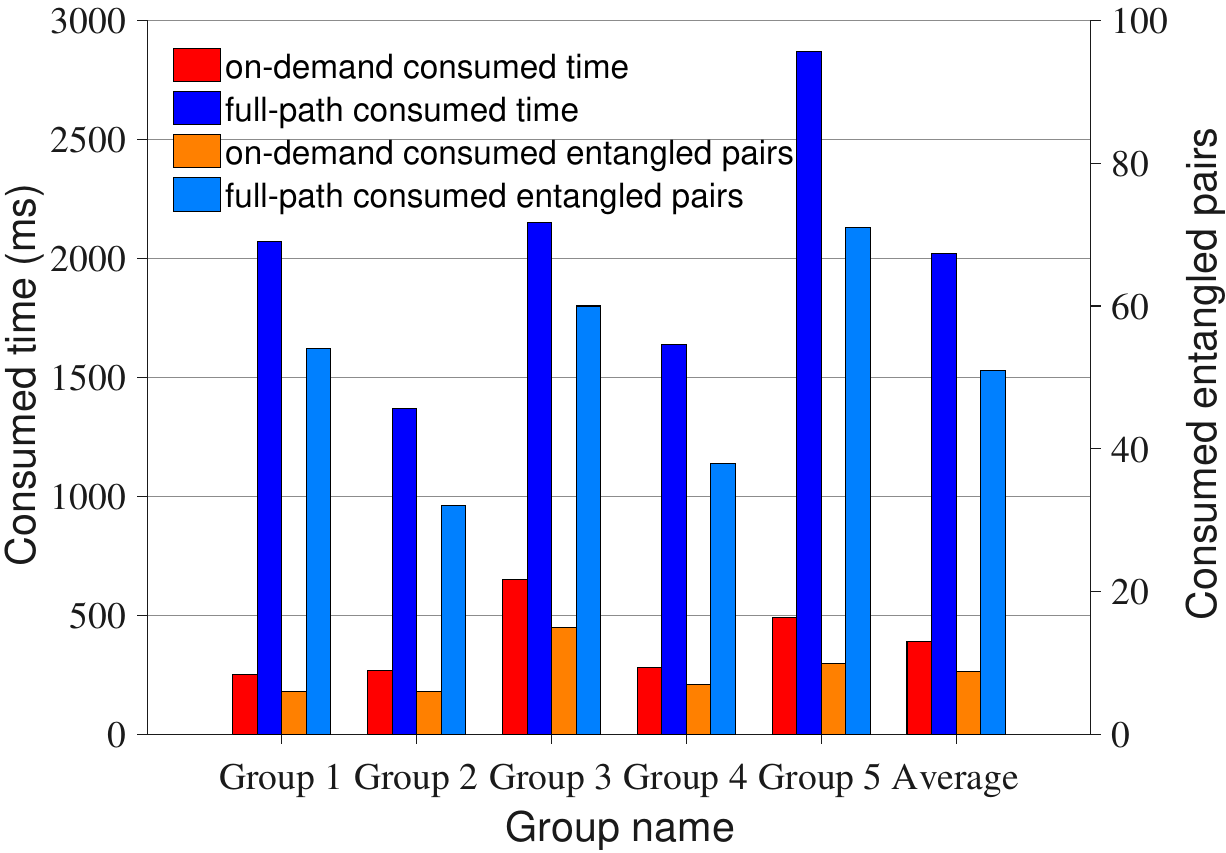}
\caption{On-demand retransimission vs. full-path retransimission. The hops are 5. The average node ES time cost is 1.4 units. The standard deviation of node ES time cost is 0.1 units.}
\label{on_demand_mechanism}
\end{minipage}
\begin{minipage}[t]{0.49\linewidth}
\centering
\includegraphics[width=1.5in]{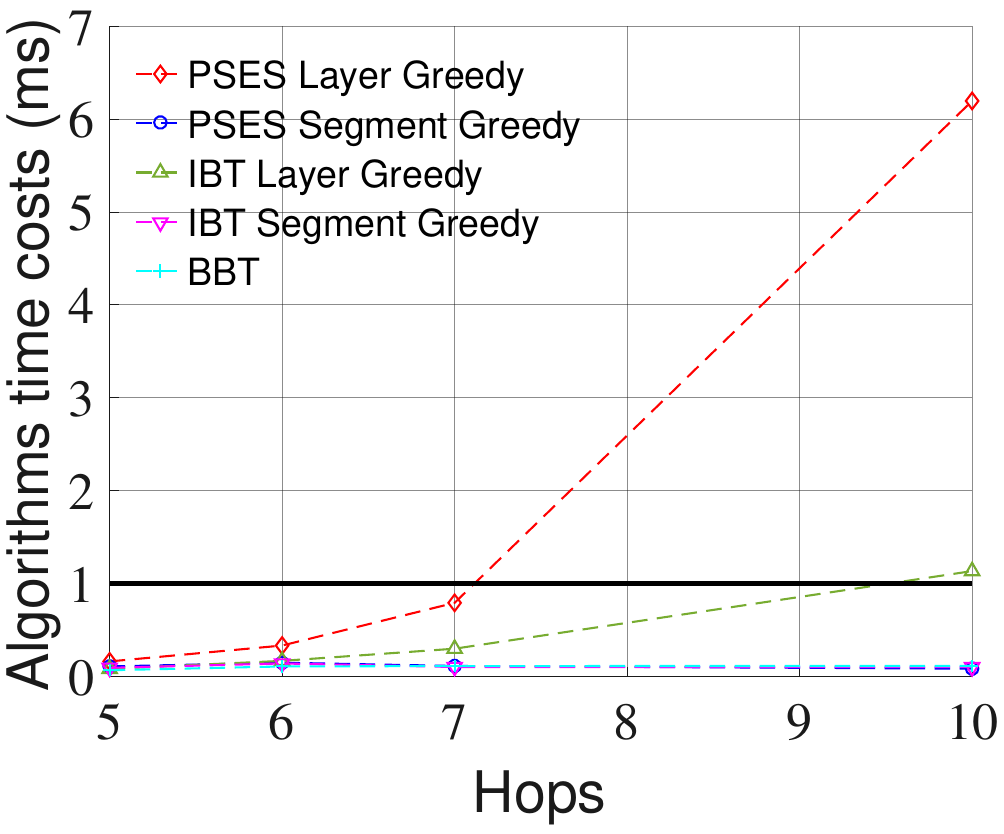}
\caption{The experiment result of time complexity}
\label{time_complexity}
\end{minipage}
\end{figure}

{\bfseries Average ES time vs. standard deviation of node ES time cost:} We note that the higher the standard deviation of node ES time cost, the greater the degree of node difference. Fig. \ref{swapping_time_vs_std}(a) shows that under different standard deviations of node ES time cost, the average ES time always has BBT $>$ IBT Segment Greedy $>$ PSES Segment Greedy $>$ IBT Layer Greedy $>$ PSES Layer Greedy. Fig. \ref{swapping_time_vs_std}(b) shows that the difference in average ES time between PSES and IBT grows with the increasing standard deviation of node ES time cost, indicating that the greater the difference between nodes, the more pronounced the advantages of PSES are. This phenomenon occurs because the greater the difference in environmental interference, the greater the difference in node ES time costs, which gives the PSES more opportunities to generate composite parent nodes, thus saving more ES time.

{\bfseries On-demand retransmission vs. full-path retransmission:} On-demand retransmission performs entanglement repreparation for the nodes related to ES failure, while full-path retransmission performs entanglement repreparation for all nodes. The red and dark blue columns in Fig. \ref{on_demand_mechanism} represent the time cost of the on-demand and full-path retransmission mechanisms, respectively. In all five groups, the time cost of the full-path retransmission is higher than that of the on-demand retransmission. The average time cost of the on-demand retransmission is only 390 ms, while that of the full-path retransmission is 2020 ms. The orange and light blue columns in Fig. \ref{on_demand_mechanism} represent the entanglement consumption of the on-demand and full-path retransmission mechanisms, respectively. In all five groups, the entanglement consumption of the full-path retransmission is higher than that of the on-demand retransmission. The average entanglement consumption of the on-demand retransmission is only 8.8, while the average entanglement consumption of the full-path retransmission is 51. We can conclude that the on-demand retransmission can reduce about 80\% of the time cost and about 80\% of entanglement consumption compared with the full-path retransmission.

{\bfseries Time complexity:} Fig. \ref{time_complexity} shows the trend of time consumption of different algorithms with hops increasing. As the hops increase, especially when the number of hops is greater than 7, the time consumption of PSES Layer Greedy and IBT Layer Greedy approaches milliseconds and has a significant upward trend. On the contrary, the time consumption of PSES Segment Greedy, IBT Segment Greedy, and BBT are at the microsecond level. The algorithm time consumption from high to low is PSES Layer Greedy $>$ IBT Layer Greedy $>$ PSES Segment Greedy $\approx$ IBT Segment Greedy $\approx$ BBT.

\subsection{Discussion}
Under the same implementation algorithm, PSES performance is always superior to IBT and BBT, regardless of hops or node ES time cost changes, because PSES introduces the composite parent node, which fully uses the idle time of parallel nodes to allow more nodes to perform ES simultaneously. The greater the difference between nodes, the more pronounced the performance advantages of PSES. On-demand retransmission can increase the long-distance entanglement generation rate by 80\% while reducing resource consumption by 80\%. When hops are less than or equal to 7 (short path), the PSES Layer Greedy, which has the best performance, should be selected because the time consumption of all strategies is in the microsecond level ($<$1ms); when hops are greater than 7 (long path), the time consumption of PSES Layer Greedy and IBT Layer Greedy both increase rapidly and approach milliseconds, so they are not suitable for parallel ES. However, the time consumption of the remaining three strategies is kept in microseconds. Therefore, PSES Segment Greedy, which performs the best among the remaining three strategies, should be selected in a long path. We should note the difference between PSES and RED\cite{dai2020optimal}. Although RED optimizes BBT well, its essence is to generate a balanced binary tree with the parent node consisting of a single repeater. PSES generates a tree-like model with the parent node consisting of multiple repeaters. The advantage of PSES is that the ES time cost of the parent node can be flexibly adjusted. Thus, there is a greater probability of getting similar ES time costs of the parallel parent nodes to save more time.

The future third-generation quantum repeater\cite{muralidharan2016optimal, munro2012quantum} may transmit qubits using the Tree Cluster States encoding\cite{borregaard2020one} instead of the ES strategy. Robustness is determined by the number of encoded qubits in the Tree Cluster States, and more encoded qubits result in more fault tolerance\cite{loke2023distributed}. However, more encoded qubits also mean consuming more resources and reducing transmission efficiency. Actually, we can use fewer encoded qubits on paths with less environmental interference. Therefore, inspired by the PSES, how to utilize the global view of hierarchical architecture to determine the number of encoded qubits of Tree Cluster States on different paths is a problem worth exploring.
\section{Conclusion and Future Work}
\label{conclusion}
This research proposes PSES to improve the efficiency of ES in long-distance quantum communication. PSES saves more time in parallel ES by introducing the composite parent node and realizes time synchronization and on-demand retransmission mechanism by using hierarchical architecture. Although we are still far from an actual quantum network, this study provides a possible ES solution for the protocol stack's network layer. Inspired by PSES, it is worth considering utilizing the advantage of hierarchical architecture to realize Tree Cluster States encoding\cite{borregaard2020one} in third-generation repeaters.

\bibliographystyle{IEEEtran}
\def\bibfont{\small}
\bibliography{reference}

\end{document}